\documentclass[conference,10pt,twocolumn,letter]{IEEEtran}

\usepackage[utf8]{inputenc} 
\usepackage[T1]{fontenc}    
\usepackage{url}            
\usepackage{graphicx}
\usepackage{nicefrac}       
\usepackage{amssymb}
\usepackage{amsmath}
\usepackage{amsthm}
\usepackage{paralist}
\usepackage{subfigure}
\usepackage{setspace}
\usepackage{booktabs}
\usepackage{algorithm}
\usepackage{microtype}
\usepackage{cite}
\usepackage{algpseudocode}
\usepackage{array}
\usepackage{balance}
\usepackage{multirow}

\usepackage{amssymb}
\usepackage{amsfonts}
\usepackage{mathrsfs}
\usepackage{xspace}
\usepackage{bm}
\usepackage{upgreek}

\newcommand{\safemath}[2]{\newcommand{#1}{\ensuremath{#2}\xspace}}

\safemath{\bma}{\mathbf{a}}
\safemath{\bmb}{\mathbf{b}}
\safemath{\bmc}{\mathbf{c}}
\safemath{\bmd}{\mathbf{d}}
\safemath{\bme}{\mathbf{e}}
\safemath{\bmf}{\mathbf{f}}
\safemath{\bmg}{\mathbf{g}}
\safemath{\bmh}{\mathbf{h}}
\safemath{\bmi}{\mathbf{i}}
\safemath{\bmj}{\mathbf{j}}
\safemath{\bmk}{\mathbf{k}}
\safemath{\bml}{\mathbf{l}}
\safemath{\bmm}{\mathbf{m}}
\safemath{\bmn}{\mathbf{n}}
\safemath{\bmo}{\mathbf{o}}
\safemath{\bmp}{\mathbf{p}}
\safemath{\bmq}{\mathbf{q}}
\safemath{\bmr}{\mathbf{r}}
\safemath{\bms}{\mathbf{s}}
\safemath{\bmt}{\mathbf{t}}
\safemath{\bmu}{\mathbf{u}}
\safemath{\bmv}{\mathbf{v}}
\safemath{\bmw}{\mathbf{w}}
\safemath{\bmx}{\mathbf{x}}
\safemath{\bmy}{\mathbf{y}}
\safemath{\bmz}{\mathbf{z}}
\safemath{\bmzero}{\mathbf{0}}
\safemath{\bmone}{\mathbf{1}}

\bmdefine{\biad}{a}
\bmdefine{\bibd}{b}
\bmdefine{\bicd}{c}
\bmdefine{\bidd}{d}
\bmdefine{\bied}{e}
\bmdefine{\bifd}{f}
\bmdefine{\bigd}{g}
\bmdefine{\bihd}{h}
\bmdefine{\biid}{i}
\bmdefine{\bijd}{j}
\bmdefine{\bikd}{k}
\bmdefine{\bild}{l}
\bmdefine{\bimd}{m}
\bmdefine{\bind}{n}
\bmdefine{\biod}{o}
\bmdefine{\bipd}{p}
\bmdefine{\biqd}{q}
\bmdefine{\bird}{r}
\bmdefine{\bisd}{s}
\bmdefine{\bitd}{t}
\bmdefine{\biud}{u}
\bmdefine{\bivd}{v}
\bmdefine{\biwd}{w}
\bmdefine{\bixd}{x}
\bmdefine{\biyd}{y}
\bmdefine{\bizd}{z}

\bmdefine{\bixid}{\xi}
\bmdefine{\bilambdad}{\lambda}
\bmdefine{\bimud}{\mu}
\bmdefine{\bithetad}{\theta}
\bmdefine{\biphid}{\phi}
\bmdefine{\bideltad}{\delta}

\safemath{\bmia}{\biad}
\safemath{\bmib}{\bibd}
\safemath{\bmic}{\bicd}
\safemath{\bmid}{\bidd}
\safemath{\bmie}{\bied}
\safemath{\bmif}{\bifd}
\safemath{\bmig}{\bigd}
\safemath{\bmih}{\bihd}
\safemath{\bmii}{\biid}
\safemath{\bmij}{\bijd}
\safemath{\bmik}{\bikd}
\safemath{\bmil}{\bild}
\safemath{\bmim}{\bimd}
\safemath{\bmin}{\bind}
\safemath{\bmio}{\biod}
\safemath{\bmip}{\bipd}
\safemath{\bmiq}{\biqd}
\safemath{\bmir}{\bird}
\safemath{\bmis}{\bisd}
\safemath{\bmit}{\bitd}
\safemath{\bmiu}{\biud}
\safemath{\bmiv}{\bivd}
\safemath{\bmiw}{\biwd}
\safemath{\bmix}{\bixd}
\safemath{\bmiy}{\biyd}
\safemath{\bmiz}{\bizd}

\safemath{\bmxi}{\bixid}
\safemath{\bmlambda}{\bilambdad}
\safemath{\bmmu}{\bimud}
\safemath{\bmtheta}{\bithetad}
\safemath{\bmphi}{\biphid}
\safemath{\bmdelta}{\bideltad}

\safemath{\bA}{\mathbf{A}}
\safemath{\bB}{\mathbf{B}}
\safemath{\bC}{\mathbf{C}}
\safemath{\bD}{\mathbf{D}}
\safemath{\bE}{\mathbf{E}}
\safemath{\bF}{\mathbf{F}}
\safemath{\bG}{\mathbf{G}}
\safemath{\bH}{\mathbf{H}}
\safemath{\bI}{\mathbf{I}}
\safemath{\bJ}{\mathbf{J}}
\safemath{\bK}{\mathbf{K}}
\safemath{\bL}{\mathbf{L}}
\safemath{\bM}{\mathbf{M}}
\safemath{\bN}{\mathbf{N}}
\safemath{\bO}{\mathbf{O}}
\safemath{\bP}{\mathbf{P}}
\safemath{\bQ}{\mathbf{Q}}
\safemath{\bR}{\mathbf{R}}
\safemath{\bS}{\mathbf{S}}
\safemath{\bT}{\mathbf{T}}
\safemath{\bU}{\mathbf{U}}
\safemath{\bV}{\mathbf{V}}
\safemath{\bW}{\mathbf{W}}
\safemath{\bX}{\mathbf{X}}
\safemath{\bY}{\mathbf{Y}}
\safemath{\bZ}{\mathbf{Z}}

\safemath{\bZero}{\mathbf{0}}
\safemath{\bOne}{\mathbf{1}}
\safemath{\bDelta}{\mathbf{\Delta}}
\safemath{\bLambda}{\mathbf{\UpLambda}}
\safemath{\bPhi}{\mathbf{\Upphi}}
\safemath{\bSigma}{\mathbf{\Upsigma}}
\safemath{\bOmega}{\mathbf{\Upomega}}
\safemath{\bTheta}{\mathbf{\Uptheta}}

\bmdefine{\biAd}{A}
\bmdefine{\biBd}{B}
\bmdefine{\biCd}{C}
\bmdefine{\biDd}{D}
\bmdefine{\biEd}{E}
\bmdefine{\biFd}{F}
\bmdefine{\biGd}{G}
\bmdefine{\biHd}{H}
\bmdefine{\biId}{I}
\bmdefine{\biJd}{J}
\bmdefine{\biKd}{K}
\bmdefine{\biLd}{L}
\bmdefine{\biMd}{M}
\bmdefine{\biOd}{N}
\bmdefine{\biPd}{O}
\bmdefine{\biQd}{P}
\bmdefine{\biRd}{R}
\bmdefine{\biSd}{S}
\bmdefine{\biTd}{T}
\bmdefine{\biUd}{U}
\bmdefine{\biVd}{V}
\bmdefine{\biWd}{W}
\bmdefine{\biXd}{X}
\bmdefine{\biYd}{Y}
\bmdefine{\biZd}{Z}

\bmdefine{\biDelta}{\Delta}
\bmdefine{\biLambda}{\Lambda}
\bmdefine{\biPhi}{\Phi}
\bmdefine{\biSigma}{\Sigma}
\bmdefine{\biOmega}{\Omega}
\bmdefine{\biTheta}{\Theta}

\safemath{\bimA}{\biAd}
\safemath{\bimB}{\biBd}
\safemath{\bimC}{\biCd}
\safemath{\bimD}{\biDd}
\safemath{\bimE}{\biEd}
\safemath{\bimF}{\biFd}
\safemath{\bimG}{\biGd}
\safemath{\bimH}{\biHd}
\safemath{\bimI}{\biId}
\safemath{\bimJ}{\biJd}
\safemath{\bimK}{\biKd}
\safemath{\bimL}{\biLd}
\safemath{\bimM}{\biMd}
\safemath{\bimN}{\biNd}
\safemath{\bimO}{\biOd}
\safemath{\bimP}{\biPd}
\safemath{\bimQ}{\biQd}
\safemath{\bimR}{\biRd}
\safemath{\bimS}{\biSd}
\safemath{\bimT}{\biTd}
\safemath{\bimU}{\biUd}
\safemath{\bimV}{\biVd}
\safemath{\bimW}{\biWd}
\safemath{\bimX}{\biXd}
\safemath{\bimY}{\biYd}
\safemath{\bimZ}{\biZd}

\safemath{\bimDelta}{\biDelta}
\safemath{\bimLambda}{\biLambda}
\safemath{\bimPhi}{\biPhi}
\safemath{\bimSigma}{\biSigma}
\safemath{\bimOmega}{\biOmega}
\safemath{\bimTheta}{\biTheta}

\safemath{\setA}{\mathcal{A}}
\safemath{\setB}{\mathcal{B}}
\safemath{\setC}{\mathcal{C}}
\safemath{\setD}{\mathcal{D}}
\safemath{\setE}{\mathcal{E}}
\safemath{\setF}{\mathcal{F}}
\safemath{\setG}{\mathcal{G}}
\safemath{\setH}{\mathcal{H}}
\safemath{\setI}{\mathcal{I}}
\safemath{\setJ}{\mathcal{J}}
\safemath{\setK}{\mathcal{K}}
\safemath{\setL}{\mathcal{L}}
\safemath{\setM}{\mathcal{M}}
\safemath{\setN}{\mathcal{N}}
\safemath{\setO}{\mathcal{O}}
\safemath{\setP}{\mathcal{P}}
\safemath{\setQ}{\mathcal{Q}}
\safemath{\setR}{\mathcal{R}}
\safemath{\setS}{\mathcal{S}}
\safemath{\setT}{\mathcal{T}}
\safemath{\setU}{\mathcal{U}}
\safemath{\setV}{\mathcal{V}}
\safemath{\setW}{\mathcal{W}}
\safemath{\setX}{\mathcal{X}}
\safemath{\setY}{\mathcal{Y}}
\safemath{\setZ}{\mathcal{Z}}
\safemath{\emptySet}{\varnothing}

\safemath{\colA}{\mathscr{A}}
\safemath{\colB}{\mathscr{B}}
\safemath{\colC}{\mathscr{C}}
\safemath{\colD}{\mathscr{D}}
\safemath{\colE}{\mathscr{E}}
\safemath{\colF}{\mathscr{F}}
\safemath{\colG}{\mathscr{G}}
\safemath{\colH}{\mathscr{H}}
\safemath{\colI}{\mathscr{I}}
\safemath{\colJ}{\mathscr{J}}
\safemath{\colK}{\mathscr{K}}
\safemath{\colL}{\mathscr{L}}
\safemath{\colM}{\mathscr{M}}
\safemath{\colN}{\mathscr{N}}
\safemath{\colO}{\mathscr{O}}
\safemath{\colP}{\mathscr{P}}
\safemath{\colQ}{\mathscr{Q}}
\safemath{\colR}{\mathscr{R}}
\safemath{\colS}{\mathscr{S}}
\safemath{\colT}{\mathscr{T}}
\safemath{\colU}{\mathscr{U}}
\safemath{\colV}{\mathscr{V}}
\safemath{\colW}{\mathscr{W}}
\safemath{\colX}{\mathscr{X}}
\safemath{\colY}{\mathscr{Y}}
\safemath{\colZ}{\mathscr{Z}}

\safemath{\opA}{\mathbb{A}}
\safemath{\opB}{\mathbb{B}}
\safemath{\opC}{\mathbb{C}}
\safemath{\opD}{\mathbb{D}}
\safemath{\opE}{\mathbb{E}}
\safemath{\opF}{\mathbb{F}}
\safemath{\opG}{\mathbb{G}}
\safemath{\opH}{\mathbb{H}}
\safemath{\opI}{\mathbb{I}}
\safemath{\opJ}{\mathbb{J}}
\safemath{\opK}{\mathbb{K}}
\safemath{\opL}{\mathbb{L}}
\safemath{\opM}{\mathbb{M}}
\safemath{\opN}{\mathbb{N}}
\safemath{\opO}{\mathbb{O}}
\safemath{\opP}{\mathbb{P}}
\safemath{\opQ}{\mathbb{Q}}
\safemath{\opR}{\mathbb{R}}
\safemath{\opS}{\mathbb{S}}
\safemath{\opT}{\mathbb{T}}
\safemath{\opU}{\mathbb{U}}
\safemath{\opV}{\mathbb{V}}
\safemath{\opW}{\mathbb{W}}
\safemath{\opX}{\mathbb{X}}
\safemath{\opY}{\mathbb{Y}}
\safemath{\opZ}{\mathbb{Z}}
\safemath{\opZero}{\mathbb{O}}
\safemath{\identityop}{\opI}

\safemath{\veca}{\bma}
\safemath{\vecb}{\bmb}
\safemath{\vecc}{\bmc}
\safemath{\vecd}{\bmd}
\safemath{\vece}{\bme}
\safemath{\vecf}{\bmf}
\safemath{\vecg}{\bmg}
\safemath{\vech}{\bmh}
\safemath{\veci}{\bmi}
\safemath{\vecj}{\bmj}
\safemath{\veck}{\bmk}
\safemath{\vecl}{\bml}
\safemath{\vecm}{\bmm}
\safemath{\vecn}{\bmn}
\safemath{\veco}{\bmo}
\safemath{\vecp}{\bmp}
\safemath{\vecq}{\bmq}
\safemath{\vecr}{\bmr}
\safemath{\vecs}{\bms}
\safemath{\vect}{\bmt}
\safemath{\vecu}{\bmu}
\safemath{\vecv}{\bmv}
\safemath{\vecw}{\bmw}
\safemath{\vecx}{\bmx}
\safemath{\vecy}{\bmy}
\safemath{\vecz}{\bmz}

\safemath{\veczero}{\bmzero}
\safemath{\vecone}{\bmone}
\safemath{\vecxi}{\bmxi}
\safemath{\veclambda}{\bmlambda}
\safemath{\vecmu}{\bmmu}
\safemath{\vectheta}{\bmtheta}
\safemath{\vecphi}{\bmphi}
\safemath{\vecdelta}{\bmdelta}

\safemath{\matA}{\bA}
\safemath{\matB}{\bB}
\safemath{\matC}{\bC}
\safemath{\matD}{\bD}
\safemath{\matE}{\bE}
\safemath{\matF}{\bF}
\safemath{\matG}{\bG}
\safemath{\matH}{\bH}
\safemath{\matI}{\bI}
\safemath{\matJ}{\bJ}
\safemath{\matK}{\bK}
\safemath{\matL}{\bL}
\safemath{\matM}{\bM}
\safemath{\matN}{\bN}
\safemath{\matO}{\bO}
\safemath{\matP}{\bP}
\safemath{\matQ}{\bQ}
\safemath{\matR}{\bR}
\safemath{\matS}{\bS}
\safemath{\matT}{\bT}
\safemath{\matU}{\bU}
\safemath{\matV}{\bV}
\safemath{\matW}{\bW}
\safemath{\matX}{\bX}
\safemath{\matY}{\bY}
\safemath{\matZ}{\bZ}
\safemath{\matzero}{\bmzero}

\safemath{\matDelta}{\bDelta}
\safemath{\matLambda}{\bLambda}
\safemath{\matPhi}{\bPhi}
\safemath{\matSigma}{\bSigma}
\safemath{\matOmega}{\bOmega}
\safemath{\matTheta}{\bTheta}

\safemath{\matidentity}{\matI}
\safemath{\matone}{\matO}

\safemath{\rnda}{A}
\safemath{\rndb}{B}
\safemath{\rndc}{C}
\safemath{\rndd}{D}
\safemath{\rnde}{E}
\safemath{\rndf}{F}
\safemath{\rndg}{G}
\safemath{\rndh}{H}
\safemath{\rndi}{I}
\safemath{\rndj}{J}
\safemath{\rndk}{K}
\safemath{\rndl}{L}
\safemath{\rndm}{M}
\safemath{\rndn}{N}
\safemath{\rndo}{O}
\safemath{\rndp}{P}
\safemath{\rndq}{Q}
\safemath{\rndr}{R}
\safemath{\rnds}{S}
\safemath{\rndt}{T}
\safemath{\rndu}{U}
\safemath{\rndv}{V}
\safemath{\rndw}{W}
\safemath{\rndx}{X}
\safemath{\rndy}{Y}
\safemath{\rndz}{Z}

\safemath{\rveca}{\bimA}
\safemath{\rvecb}{\bimB}
\safemath{\rvecc}{\bimC}
\safemath{\rvecd}{\bimD}
\safemath{\rvece}{\bimE}
\safemath{\rvecf}{\bimF}
\safemath{\rvecg}{\bimG}
\safemath{\rvech}{\bimH}
\safemath{\rveci}{\bimI}
\safemath{\rvecj}{\bimJ}
\safemath{\rveck}{\bimK}
\safemath{\rvecl}{\bimL}
\safemath{\rvecm}{\bimM}
\safemath{\rvecn}{\bimN}
\safemath{\rveco}{\bomO}
\safemath{\rvecp}{\bimP}
\safemath{\rvecq}{\bimQ}
\safemath{\rvecr}{\bimR}
\safemath{\rvecs}{\bimS}
\safemath{\rvect}{\bimT}
\safemath{\rvecu}{\bimU}
\safemath{\rvecv}{\bimV}
\safemath{\rvecw}{\bimW}
\safemath{\rvecx}{\bimX}
\safemath{\rvecy}{\bimY}
\safemath{\rvecz}{\bimZ}

\safemath{\rvecxi}{\bmxi}
\safemath{\rveclambda}{\bmlambda}
\safemath{\rvecmu}{\bmmu}
\safemath{\rvectheta}{\bmtheta}
\safemath{\rvecphi}{\bmphi}

\safemath{\rmatA}{\bimA}
\safemath{\rmatB}{\bimB}
\safemath{\rmatC}{\bimC}
\safemath{\rmatD}{\bimD}
\safemath{\rmatE}{\bimE}
\safemath{\rmatF}{\bimF}
\safemath{\rmatG}{\bimG}
\safemath{\rmatH}{\bimH}
\safemath{\rmatI}{\bimI}
\safemath{\rmatJ}{\bimJ}
\safemath{\rmatK}{\bimK}
\safemath{\rmatL}{\bimL}
\safemath{\rmatM}{\bimM}
\safemath{\rmatN}{\bimN}
\safemath{\rmatO}{\bimO}
\safemath{\rmatP}{\bimP}
\safemath{\rmatQ}{\bimQ}
\safemath{\rmatR}{\bimR}
\safemath{\rmatS}{\bimS}
\safemath{\rmatT}{\bimT}
\safemath{\rmatU}{\bimU}
\safemath{\rmatV}{\bimV}
\safemath{\rmatW}{\bimW}
\safemath{\rmatX}{\bimX}
\safemath{\rmatY}{\bimY}
\safemath{\rmatZ}{\bimZ}

\safemath{\rmatDelta}{\bimDelta}
\safemath{\rmatLambda}{\bimLambda}
\safemath{\rmatPhi}{\bimPhi}
\safemath{\rmatSigma}{\bimSigma}
\safemath{\rmatOmega}{\bimOmega}
\safemath{\rmatTheta}{\bimTheta}

\usepackage{amssymb}
\usepackage{amsfonts}
\usepackage{mathrsfs}
\usepackage{xspace}
\usepackage{bm}
\usepackage{fancyref}
\usepackage{textcomp}

\usepackage{multirow}
\usepackage{stmaryrd}

\newenvironment{textbmatrix}{	\setlength{\arraycolsep}{2.5pt}%
								\big[\begin{matrix}}{\end{matrix}\big]%
								\raisebox{0.08ex}{\vphantom{M}}}

\def\be{\begin{equation}}
\def\ee{\end{equation}}
\def\een{\nonumber \end{equation}}
\def\mat{\begin{bmatrix}}
\def\emat{\end{bmatrix}}
\def\btm{\begin{textbmatrix}}
\def\etm{\end{textbmatrix}}

\def\ba#1\ea{\begin{align}#1\end{align}}
\def\bas#1\eas{\begin{align*}#1\end{align*}}
\def\bs#1\es{\begin{split}#1\end{split}} 
\def\bg#1\eg{\begin{gather}#1\end{gather}}
\def\bml#1\eml{\begin{multline}#1\end{multline}}
\def\bi#1\ei{\begin{itemize}#1\end{itemize}}

\newcommand{\lefto}{\mathopen{}\left}

\DeclareMathOperator{\tr}{tr}				
\DeclareMathOperator{\diag}{diag}			
\DeclareMathOperator{\Exop}{\opE}			
\DeclareMathOperator{\Varop}{\mathrm{Var}}
\DeclareMathOperator{\Covop}{\mathrm{Cov}}

\newcommand{\Ex}[2]{\ensuremath{\Exop_{#1}\lefto[#2\right]}} 	
\newcommand{\Var}[1]{\ensuremath{\Varop\lefto[#1\right]}} 
\newcommand{\Cov}[1]{\ensuremath{\Covop\lefto[#1\right]}} 

\safemath{\dirac}{\delta}					
\safemath{\krond}{\dirac}					

\safemath{\upto}{\uparrow}
\safemath{\downto}{\downarrow}
\safemath{\iu}{j}							
\safemath{\ev}{\lambda}						
\safemath{\hilseqspace}{l^{2}}				
\newcommand{\banachfunspace}[1]{\setL^{#1}}	
\safemath{\hilfunspace}{\banachfunspace{2}}	

\safemath{\SNR}{\textsf{SNR}} 				
\safemath{\PAR}{\textsf{PAR}} 				
\safemath{\No}{N_0}							
\safemath{\Es}{E_s}							
\safemath{\Eb}{E_b}							
\safemath{\EbNo}{\frac{\Eb}{\No}}
\safemath{\EsNo}{\frac{\Es}{\No}}

\DeclareMathOperator{\CHop}{\ensuremath{\opH}} 
\safemath{\tvir}{\rndh_{\CHop}}				
\safemath{\tvtf}{\rndl_{\CHop}}				
\safemath{\spf}{\rnds_{\CHop}}				
\safemath{\bff}{H_{\CHop}}					

\safemath{\ircf}{r_{h}}						
\safemath{\tftvcf}{r_{s}}					
\safemath{\tfcf}{r_{l}}						
\safemath{\bfcf}{r_{H}}						

\safemath{\tcorr}{c_h}						
\safemath{\scf}{c_{s}}						
\safemath{\tfcorr}{c_{l}}					
\safemath{\fcorr}{c_{H}}						

\safemath{\mi}{I}							
\safemath{\capacity}{C}						

\safemath{\normal}{\mathcal{N}}			
\safemath{\jpg}{\mathcal{CN}}			
\safemath{\mchain}{\leftrightarrow}		

\safemath{\dB}{\,\mathrm{dB}}
\safemath{\dBm}{\,\mathrm{dBm}}
\safemath{\Hz}{\,\mathrm{Hz}}
\safemath{\kHz}{\,\mathrm{kHz}}
\safemath{\MHz}{\,\mathrm{MHz}}
\safemath{\GHz}{\,\mathrm{GHz}}
\safemath{\s}{\,\mathrm{s}}
\safemath{\ms}{\,\mathrm{ms}}
\safemath{\mus}{\,\mathrm{\text{\textmu}s}}
\safemath{\ns}{\,\mathrm{ns}}
\safemath{\ps}{\,\mathrm{ps}}
\safemath{\meter}{\,\mathrm{m}}
\safemath{\mm}{\,\mathrm{mm}}
\safemath{\cm}{\,\mathrm{cm}}
\safemath{\m}{\,\mathrm{m}}
\safemath{\W}{\,\mathrm{W}}
\safemath{\mW}{\, \mathrm{mW}}
\safemath{\J}{\,\mathrm{J}}
\safemath{\K}{\,\mathrm{K}}
\safemath{\bit}{\,\mathrm{bit}}
\safemath{\nat}{\,\mathrm{nat}}

\safemath{\define}{\triangleq}			

\safemath{\equivalent}{\sim}
\safemath{\distas}{\sim}					
\safemath{\sdiff}{\Delta}				

\safemath{\reals}{\mathbb{R}}
\safemath{\positivereals}{\reals_{+}}
\safemath{\integers}{\mathbb{Z}}
\safemath{\posint}{\integers_{+}}
\safemath{\naturals}{\mathbb{N}}
\safemath{\posnaturals}{\naturals_{+}}
\safemath{\complexset}{\mathbb{C}}
\safemath{\rationals}{\mathbb{Q}}

\newcommand*{\fancyrefapplabelprefix}{app}		
\newcommand*{\fancyrefthmlabelprefix}{thm}		
\newcommand*{\fancyreflemlabelprefix}{lem}		
\newcommand*{\fancyrefcorlabelprefix}{cor}		
\newcommand*{\fancyrefdeflabelprefix}{def}		
\newcommand*{\fancyrefproplabelprefix}{prop}	
\newcommand*{\fancyrefobslabelprefix}{obs}		
\newcommand*{\fancyrefalglabelprefix}{alg}		
\newcommand*{\fancyrefasmlabelprefix}{asm}	    
\newcommand*{\fancyrefasmslabelprefix}{asms}	    
\newcommand*{\fancyreftbllabelprefix}{tbl}	    

\frefformat{vario}{\fancyrefseclabelprefix}{Section~#1}
\frefformat{vario}{\fancyrefthmlabelprefix}{Theorem~#1}
\frefformat{vario}{\fancyreflemlabelprefix}{Lemma~#1}
\frefformat{vario}{\fancyrefcorlabelprefix}{Corollary~#1}
\frefformat{vario}{\fancyrefdeflabelprefix}{Definition~#1}
\frefformat{vario}{\fancyrefobslabelprefix}{Observation~#1}
\frefformat{vario}{\fancyrefasmlabelprefix}{Assumption~#1}
\frefformat{vario}{\fancyrefasmslabelprefix}{Assumptions~#1}
\frefformat{vario}{\fancyreffiglabelprefix}{Figure~#1}
\frefformat{vario}{\fancyrefapplabelprefix}{Appendix~#1} 
\frefformat{vario}{\fancyrefproplabelprefix}{Proposition~#1}
\frefformat{vario}{\fancyrefalglabelprefix}{Algorithm~#1}
\frefformat{vario}{\fancyrefeqlabelprefix}{(#1)}
\frefformat{vario}{\fancyreftbllabelprefix}{Table~#1}
\newtheorem{thm}{Theorem}

\newtheorem{lem}[thm]{Lemma} 

\newtheorem{rem}{Remark}

\newtheorem{asms}{Assumptions}

\safemath{\dictab}{[\,\dicta\,\,\dictb\,]}

\safemath{\ysig}{\bmy}
\safemath{\ysighat}{\hat{\ysig}}
\safemath{\ysigdim}{M}
\safemath{\xsig}{\bmx}
\safemath{\xsigdim}{N}
\safemath{\nx}{n_x}
\safemath{\zsig}{\bmz}
\safemath{\zsigdim}{\ysigdim}
\safemath{\rsig}{\bmr}
\safemath{\Adict}{\bA}
\safemath{\Adicttilde}{\widetilde{\Adict}}
\safemath{\Adictdim}{\outputdim\times\xsigdim}
\safemath{\avec}{\bma}
\safemath{\avectilde}{\tilde{\avec}}
\safemath{\Bdict}{\bB}
\safemath{\Bdicttilde}{\widetilde{\Bdict}}
\safemath{\Cdict}{\bC}
\safemath{\cvec}{\bmc}
\safemath{\Ddict}{\bD}
\safemath{\Ddictdim}{\ysigdim\times\xsigdim}
\safemath{\dvec}{\bmd}
\safemath{\Ddicttilde}{\widetilde{\bD}}
\safemath{\Bonb}{\bB}
\safemath{\bvec}{\bmb}
\safemath{\Bonbdim}{\ysigdim\times\ysigdim}
\safemath{\noise}{\bmn}
\safemath{\noisedim}{\ysigim}
\safemath{\err}{\bme}
\safemath{\errdim}{\ysigdim}
\safemath{\errset}{\setE}
\safemath{\nerr}{n_e}
\safemath{\delop}{\bP_\errset}
\safemath{\delopc}{\bP_{{\errset}^c}}

\safemath{\cplxi}{\imath}
\safemath{\cplxj}{\jmath}

\safemath{\dict}{\matD}
\safemath{\inputdim}{N}		
\safemath{\outputdim}{M}		
\safemath{\sparsity}{S}	
\safemath{\inputdimA}{{N_a}}	
\safemath{\inputdimB}{{N_b}}	
\safemath{\elemA}{{n_a}}	
\safemath{\elemB}{{n_b}}	
\safemath{\resA}{\matR_a}	
\safemath{\resB}{\matR_b}	
\safemath{\subD}{\matS} 
\safemath{\subA}{\matS_a} 
\safemath{\subB}{\matS_b} 
\safemath{\dicta}{\matA} 	
\safemath{\dictb}{\matB} 	
\safemath{\hollowS}{H}
\safemath{\hollowA}{H_a}
\safemath{\hollowB}{H_b}
\safemath{\cross}{Z}
\safemath{\coh}{\mu_d}			
\safemath{\coha}{\mu_a}			
\safemath{\cohb}{\mu_b}			
\safemath{\mubs}{\nu}	
\safemath{\cohm}{\mu_m} 
\safemath{\dictset}{\setD}	
\safemath{\dictsetp}{\dictset(\coh,\coha,\cohb)}	
\safemath{\dictsetgen}{\dictset_\text{gen}}
\safemath{\dictsetgenp}{\dictsetgen(\coh)}
\safemath{\dictsetonb}{\dictset_\text{onb}}
\safemath{\dictsetonbp}{\dictsetonb(\coh)}

\safemath{\leftside}{U}
\safemath{\rightsideA}{R_a}
\safemath{\rightsideB}{R_b}

\safemath{\indexS}{\setI_S} 

\safemath{\na}{n_a}			
\safemath{\nb}{n_b}			
\safemath{\coeffa}{p_i}	
\safemath{\coeffb}{q_j}	
\safemath{\seta}{\setP}		
\safemath{\setb}{\setQ}     
\safemath{\setw}{\setW}	
\safemath{\setz}{\setZ}	
\safemath{\cola}{\veca}		
\safemath{\colb}{\vecb}		
\safemath{\cold}{\vecd}		
\safemath{\inputvec}{\vecx} 	
\safemath{\error}{\vece}	
\safemath{\noiseout}{\vecz} 	
\safemath{\inputvecel}{x}
\safemath{\inputveca}{\vecx_a}
\safemath{\inputvecb}{\vecx_b}
\safemath{\outputvec}{\vecy}	
\safemath{\lambdamin}{\lambda_{\mathrm{min}}}

\safemath{\elltwo}{\ell_2}
\safemath{\ellone}{\ell_1}
\safemath{\ellzero}{\ell_0}
\safemath{\ellinf}{\ell_\infty}
\safemath{\ellinftilde}{\ell_{\widetilde\infty}}
\safemath{\licard}{Z(\coh,\coha,\cohb)}
\safemath{\xsol}{\hat{x}}
\safemath{\xbord}{x_b}		
\safemath{\xstat}{x_s}		
\safemath{\xstatLone}{\tilde{x}_s}
\safemath{\order}{\mathcal{O}} 
\safemath{\scales}{\Theta} 
\safemath{\ones}{\mathbf{1}} 
\safemath{\zeroes}{\mathbf{0}} 
\safemath{\thlone}{\kappa(\coh,\cohb)} 
\safemath{\constoneA}{\delta} 
\safemath{\constoneB}{\epsilon} 
\safemath{\nlarge}{L}				   
\safemath{\sumlarge}{S_\nlarge}
\safemath{\maxlarger}{P_\nlarge}	   
\safemath{\Pzero}{\textrm{P0}}	
\safemath{\Pone}{\textrm{P1}}
\safemath{\vecfir}{\vecw}			 
\safemath{\vecsec}{\vecz}
\safemath{\elvecfir}{w}              
\safemath{\elvecsec}{z}				 
\safemath{\nlargefir}{n}
\safemath{\normout}{\gamma}
\safemath{\auxfun}{h}
\safemath{\supp}{\textrm{supp}}

\safemath{\indexa}{\ell}
\safemath{\indexb}{r}
\safemath{\indexc}{i}
\safemath{\indexd}{j}

\safemath{\project}{P}

\allowdisplaybreaks

\allowdisplaybreaks 
\IEEEoverridecommandlockouts

\usepackage{color}

\safemath{\Herm}{\textnormal{H}}
\safemath{\Tran}{\textnormal{T}}

\begin{document}

\title{PhaseLin: Linear Phase Retrieval}

\author{
\IEEEauthorblockN{Ramina Ghods$^{1}$, Andrew S. Lan$^{2}$, Tom Goldstein$^{3}$, and Christoph Studer$^{1}$}\\ \vspace{-0.1cm}
\IEEEauthorblockA{$^\text{1}$Cornell University, Ithaca, NY;  rg548@cornell.edu, studer@cornell.edu} 
\IEEEauthorblockA{$^\text{2}$Princeton University, Princeton, NJ; {andrew.lan@princeton.edu}} 
\IEEEauthorblockA{$^\text{3}$University of Maryland, College Park, MD; {tomg@cs.umd.edu}} 
\thanks{RG and CS were supported in part by Xilinx Inc.~and by the US National Science Foundation (NSF) under grants ECCS-1408006, CCF-1535897, CAREER CCF-1652065, and CNS-1717559. TG was supported in part by the US NSF under grant CCF-1535902, the US Office of Naval Research under grant N00014-17-1-2078, and by the Sloan Foundation.}
\thanks{RG and CS would like to thank S. Jacobsson, G. Durisi, and O. Tirkkonen for discussions on linearized phase retrieval methods.}
}

\maketitle

\begin{abstract}
Phase retrieval deals with the recovery of complex- or real-valued signals from magnitude measurements.
As shown recently, the method PhaseMax enables phase retrieval via convex optimization and without lifting the problem to a higher dimension.
To succeed, PhaseMax requires an initial guess of the solution, which can be calculated via spectral initializers. 
In this paper, we show that with the availability of an initial guess,  phase retrieval can be carried out with an ever simpler, linear procedure.
Our algorithm, called PhaseLin, is the linear estimator that minimizes the mean squared error (MSE) when applied to the magnitude measurements. The linear nature of PhaseLin enables an exact and nonasymptotic MSE analysis for arbitrary  measurement matrices. 
We furthermore demonstrate that by iteratively using PhaseLin, one arrives at an efficient phase retrieval  algorithm that performs on par with existing convex and nonconvex methods on synthetic and real-world data.
\end{abstract}




\section{Introduction}
\label{sec:introduction}

Phase retrieval recovers the $N$-dimensional signal vector $\bmx\in\setH^N$, with~$\setH$ being either the set of real ($\reals$) or complex ($\complexset$) numbers, from the following nonlinear measurement process:  
\begin{align} \label{eq:measurementmodel}
\bmy = |\bA\bmx+\bmn^z|^2+\bmn^y.
\end{align}
Here, the measurement vector $\bmy\in\reals^M$ contains $M$ real-valued and  phase-less observations, the  absolute square function $|\cdot|^2$ operates element-wise on vectors, \mbox{$\bA\in\setH^{M\times N}$} is a known measurement matrix, and the vectors~$\bmn^z\in\setH^N$ and~$\bmn^y\in\reals^M$ model signal and measurement noise, respectively. 
In what follows, we assume a deterministic (and known) measurement matrix $\bA$, but randomness in the signal $\bmx$ to be estimated as well as the two noise sources $\bmn^z$ and $\bmn^y$.


\subsection{Relevant Prior Art}
Phase retrieval has  been studied extensively over the last few decades~\cite{gerchberg1972practical,fienup1982phase} as it finds use in numerous applications, including X-ray crystallography \cite{harrison1993phase,miao2008extending,pfeiffer2006phase}, microscopy \cite{kou2010transport,faulkner2004movable}, and imaging \cite{holloway2016}.
In its original form, the phase retrieval problem is nonconvex and was solved traditionally using alternating projection methods, such as the Gerchberg-Saxton~\cite{gerchberg1972practical} or Fienup~\cite{fienup1982phase} methods. 
During the  last few years, it has been shown that lifting the phase retrieval problem to a higher-dimensional space enables the use of semidefinite programming~\cite{candes2013phaselift,candes2014solving,candes2015phase}, which led to a revival of phase retrieval research. 
While lifting-based phase retrieval methods provide strong theoretical guarantees, their computational complexity and storage requirements quickly become prohibitive for high-dimensional problems (e.g., for the recovery of high-resolution images).
To perform phase retrieval for high-dimensional problems with methods that provide theoretical performance guarantees, a number of nonconvex methods have been proposed and analyzed within the last years; see, e.g.,~\cite{netrapalli2013phase,schniter2015compressive,candes2015wirtinger,chen2015solving,wang2016solving,wei2015solving,zeng2017coordinate,lu2017phase,mondelli2017fundamental}. All these methods rely on an accurate initial guess of the true signal to be recovered, which can be obtained by means of so-called spectral initializers~\cite{netrapalli2013phase,chen2015solving,lu2017phase,mondelli2017fundamental,chen2015phase,wang2016solving,wang2017solving}.
More recently, it has been shown in \cite{bahmani2017phase,goldstein2016phasemax,phasemax,dhifallah2017phase,dhifallah2017fundamental} that given such initial guesses, one can perform phase retrieval via the convex program PhaseMax, which avoids lifting and provides accurate performance guarantees.

\subsection{Contributions}
In this paper, we show that the availability of an initial guess enables the use of \emph{linear} estimators to perform phase retrieval.
Concretely, we propose a novel mean squared error (MSE)-optimal linear phase retrieval method we call \emph{PhaseLin}. 
Our method provides exact and nonasymptotic expressions of the recovery MSE and is suitable for scenarios in which the measurement matrix is finite-dimensional, deterministic, and (possibly) structured---these aspects are in stark contrast to most existing theoretical results that are either of asymptotic nature and/or require randomness in the measurement matrix. 
We furthermore show that by iteratively using PhaseLin, one arrives at a phase retrieval algorithm that performs on par with existing methods on synthetic and real-world data.

\subsection{Notation}

%
Column vectors and matrices are denoted with lower- and upper-case bold letters, respectively.  For a matrix $\bA$, its transpose, Hermitian transpose, and (entry-wise) complex conjugate are denoted respectively by $\bA^\Tran$, $\bA^\Herm$, and $\bA^*$. The entry on the $m$th row and $n$th column of $\bA$ is denoted as $[\bA]_{m,n}$, and $\bma_m$ denotes the $m$th column vector.
For a vector~$\bma$, its $k$th entry is denoted by $a_k$; the $\ell_2$-norm norm is denoted by~$\|\bma\|_2$.  The (entry-wise) Hadamard product and the trace operator are denoted by $\odot$ and $\tr(\cdot)$, respectively.
The notation $\diag(\bma)$ stands for the square matrix with the vector $\bma$ on its main diagonal; $\diag(\bA)$ denotes the column vector comprising the diagonal elements of the matrix $\bA$.
The operators $\Re(\cdot)$ and $\Im(\cdot)$ extract the real and imaginary parts of a complex-valued number, respectively; for a complex-valued vector $\bma$, we use $\bma_\mathcal{R}$ and $\bma_\mathcal{I}$ to denote the real and imaginary parts.  
Expectation with respect to the random vector $\bmx$ is denoted by~$\Ex{\bmx}{\cdot}$.

\section{Main Results}\label{sec:phaselin}
We now present PhaseLin.  We separately provide results for the recovery of real-valued and complex-valued signals. We then provide an exact expression for the recovery MSE of PhaseLin.
We finally show how one can iteratively use PhaseLin to arrive at a powerful phase retrieval algorithm.

\subsection{The Real Case: PhaseLin-$\reals$}\label{sec:realcase}
We first focus on the case where the signal to be recovered $\bmx\in\reals^N$ and the measurement matrix $\bA\in\reals^{M\times N}$ are both real-valued. We need the following assumptions.

\begin{asms} \label{asms:realPhaseLin}
We assume that the signal noise vector $\bmn^z$ is zero-mean Gaussian with covariance matrix $\bC_{\bmn^z}$, i.e., $\bmn^z \sim \setN(\bZero,\bC_{\bmn^z})$; the measurement noise vector $\bmn^y$ is Gaussian with mean $\bar\bmn^y$ and covariance matrix $\bC_{\bmn^y}$, i.e.,  $\bmn^y \sim \setN(\bar\bmn^y,\bC_{\bmn^y})$.
For the signal vector $\bmx\in\reals^N$, we assume that we have an initial guess $\bar\bmx\in\reals^N$ (e.g., obtained from a spectral initializer).  We furthermore assume that the true signal can be written as $\bmx=\bar\bmx+\bme$, where the error vector $\bme$ denotes the difference between the initial guess $\bar\bmx$ and the true signal $\bmx$.  We assume that the error vector follows a zero-mean Gaussian distribution with covariance  $\bC_\bme$, i.e., $\bme\sim\setN(\bZero,\bC_\bme)$. 
\end{asms}

With these assumptions, we can derive PhaseLin-$\reals$; the proof of the following result is given in \fref{app:realPhaseLin}.

\begin{thm}[PhaseLin-$\reals$] \label{thm:realPhaseLin}
Under \fref{asms:realPhaseLin}, the linear estimate $\hat\bmx$ that minimizes the recovery MSE defined as
\begin{align} \label{eq:MSE}
\textit{MSE} = \Ex{\,\bme,\bmn^z,\bmn^y}{\|\hat\bmx-\bmx\|^2_2}
\end{align}
is given by
\begin{align*}
\hat\bmx = \bC_{\bmx,\bmy} \bmv + \bar\bmx \quad \text{with} \quad \bC_\bmy\bmv= \bmy-\bar\bmy,
\end{align*}
where $\bar\bmx$ is an initial guess and
\begin{align*}
\bar \bmy & = \diag(\bC_\bmz)+|\bar{\bmz}|^2+\bar\bmn^y \\
\bC_{\bmx,\bmy} & = 2 \bC_{\bme}\bA^\Tran\diag(\bar{\bmz})  \\
\bC_{\bmy} & =  (4\bar{\bmz}\bar{\bmz}^\Tran + 2\bC_{\bmz})  \odot  \bC_{\bmz} \! + \!\bC_{\bmn^y},
\end{align*}
with
$\bar{\bmz}=\bA\bar{\bmx}$ and $\bC_{\bmz}=\bA \bC_{\bme} \bA^\Tran+\bC_{\bmn^z}$.
\end{thm}

The above result describes the linear estimate~$\hat\bmx$ of the signal~$\bmx$ to be recovered that minimizes the recovery MSE, given the phase-less measurements in $\bmy$ and an initial guess $\bar\bmx$. As shown in \fref{lem:MSE}, we are able to provide a closed form expression for the recovery MSE of PhaseLin-$\reals$. 

\begin{rem}
If the initial guess $\bar\bmx$ is zero, then the quantity $\bar\bmz$ is zero. 
In this situation, the matrix $\bC_{\bmx,\bmy}$ is zero and, hence, the obtained estimate $\hat\bmx$ is zero as well. 
Clearly, for such an initial guess, PhaseLin fails at estimating the signal $\bmx$. 
By using spectral initializers to set the mean~$\bar\bmx$, such as the ones proposed in \cite{netrapalli2013phase,chen2015solving,lu2017phase,mondelli2017fundamental,chen2015phase,wang2016solving,wang2017solving}, PhaseLin performs well in practice; see \fref{sec:results} for simulation results.
\end{rem}

\subsection{The Complex Case: PhaseLin-$\complexset$}\label{sec:complexcase}
We now focus on the case where the signal to be recovered $\bmx\in\complexset^N$ and the measurement matrix $\bA\in\complexset^{M\times N}$ are both complex-valued. The measurements, however, remain real-valued. We need the following assumptions.

\begin{asms} \label{asms:complexPhaseLin}
We assume that the signal noise vector $\bmn^z$ is circularly symmetric complex Gaussian with covariance matrix~$\bC_{\bmn^z}$, i.e., $\bmn^z \sim \setC\setN(\bZero,\bC_{\bmn^z})$.
%
%
We assume that the error vector is also circularly symmetric complex Gaussian with covariance matrix~$\bC_\bme$, i.e., $\bme\sim\setC \setN(\bZero,\bC_\bme)$.  
The remaining assumptions are identical to those in \fref{asms:realPhaseLin}.  
%
\end{asms}

We can now derive PhaseLin-$\complexset$; the proof of the following result is given in \fref{app:complexPhaseLin}.

\begin{thm}[PhaseLin-$\complexset$] \label{thm:complexPhaseLin}
	Under \fref{asms:complexPhaseLin}, the linear estimate $\hat\bmx$ that minimizes the MSE in \fref{eq:MSE} 
	is given by
	\begin{align*}
	\hat\bmx = \bC_{\bmx,\bmy} \bmv + \bar\bmx \quad \text{with} \quad \bC_\bmy\bmv= \bmy-\bar\bmy,
	\end{align*}
	where $\bar\bmx$ is an initial guess and
	\begin{align*}
	\bar \bmy &  = \diag(\bC_\bmz)+|\bar{\bmz}|^2+\bar\bmn^y \\
	\bC_{\bmx,\bmy} & = \bC_{\bme}\bA^\Herm \diag(\bar{\bmz})  \\	
		\bC_{\bmy} & = 2\Re\!\left\{\left(\bar{\bmz}\bar{\bmz}^\Herm\right) \odot \bC_{\bmz} ^* \right\}+\bC_{\bmz} \odot \bC_{\bmz}^*+\bC_{\bmn^y},
	\end{align*}
	with
	$\bar{\bmz}=\bA\bar{\bmx}$ and $\bC_{\bmz}=\bA \bC_{\bme} \bA^\Herm+\bC_{\bmn^z}$.
	
\end{thm}

As for PhaseLin-$\complexset$, the above result describes the linear estimate~$\hat\bmx$ of the signal $\bmx$  that minimizes the recovery MSE given the phase-less measurements in $\bmy$ and an initial guess $\bar\bmx$. 

\setlength{\textfloatsep}{10pt}
\begin{figure*}[tp]
	\centering
	\subfigure[Original]{
		\includegraphics[scale=0.7]{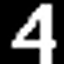}
	} \hspace{0.0cm}
	\subfigure[PhaseLin-$\complexset$]{
		\includegraphics[scale=0.69]{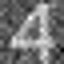}
	} \hspace{0.0cm}
	\subfigure[WF]{
		\includegraphics[scale=0.69]{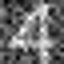}
	} \hspace{0.0cm}
	\subfigure[RAF]{
		\includegraphics[scale=0.69]{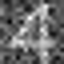}
	} \hspace{0.0cm}
	\subfigure[Fienup]{
		\includegraphics[scale=0.69]{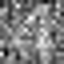}
	} \hspace{0.0cm}
	\subfigure[PhaseMax]{
		\includegraphics[scale=0.69]{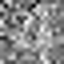}
	} \hspace{0.0cm}
	\subfigure[PhaseLamp]{
		\includegraphics[scale=0.69]{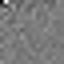}
	} \hspace{0.0cm}
	\subfigure[GS]{
		\includegraphics[scale=0.69]{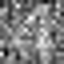}
	} \hspace{0.0cm}
	\subfigure[PhaseLift]{
		\includegraphics[scale=0.69]{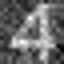}
	}
	\vspace{-0.15cm}
	
\vspace{2mm}

	\vspace{-0.2cm}
	\centering
	\subfigure[Original]{
		\includegraphics[scale=0.69]{Figs/TM16-5-original.png}
	} \hspace{0.0cm}
	\subfigure[PhaseLin-$\complexset$]{
		\includegraphics[scale=0.69]{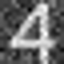}
	} \hspace{0.0cm}
	\subfigure[WF]{
		\includegraphics[scale=0.69]{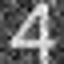}
	} \hspace{0.0cm}
	\subfigure[RAF]{
		\includegraphics[scale=0.69]{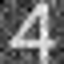}
	} \hspace{0.0cm}
	\subfigure[Fienup]{
		\includegraphics[scale=0.69]{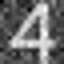}
	} \hspace{0.0cm}
	\subfigure[PhaseMax]{
		\includegraphics[scale=0.69]{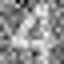}
	} \hspace{0.0cm}
	\subfigure[PhaseLamp]{
		\includegraphics[scale=0.69]{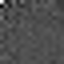}
	} \hspace{0.0cm}
	\subfigure[GS]{
		\includegraphics[scale=0.69]{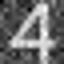}
	} \hspace{0.0cm}
	\subfigure[PhaseLift]{
		\includegraphics[scale=0.69]{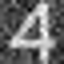}
	}
	\vspace{-0.15cm}
	\caption{ (Top) Reconstructions of a $16 \times 16$-pixel image captured through a scattering medium (real-world dataset provided in \cite{metzler2017coherent}, reconstructions using PhasePack~\cite{chandra2017phasepack}) with $ M=3  N$ measurements using different phase retrieval methods.  (Bottom) The same experiment but with $ M=9  N$ measurements. See \fref{tbl:TM16} for the  N-MSE values and runtimes of each phase retrieval algorithm.}	
	\vspace{-0.2cm}
	\label{fig:TM16}
\end{figure*}

\subsection{Exact Expression for the Recovery MSE}

For both PhaseLin-$\reals$ and PhaseLin-$\complexset$, 
the following result provides an exact and nonasymptotic expression of the recovery MSE;
the proof is given in \fref{app:MSE}.

\begin{lem}[MSE of PhaseLin] \label{lem:MSE}
	Let either \fref{asms:realPhaseLin} or \fref{asms:complexPhaseLin} hold. Furthermore, assume that $\bC_\bmy$ is full rank. Then, the recovery MSE is given by
	\begin{align*}
	\textit{MSE} = \tr(\bC_{\bme}-\bC_{\bmx,\bmy} \bC_{\bmy}^{-1} \bC_{\bmx,\bmy}^\Herm).
	\end{align*} 
\end{lem}

We emphasize that most existing phase retrieval methods either provide theoretical results that are exact in the asymptotic regime or provide upper bounds on the recovery error. In addition, virtually all existing theoretical results require randomness in the measurement matrix $\bA$. In contrast, \fref{lem:MSE} assumes randomness in the signal to be recovered and the noise sources, is nonasymptotic and exact, and holds for arbitrary and deterministic measurement matrices. 

\begin{rem}
\fref{lem:MSE} requires the matrix $\bC_\bmy$ to be full rank. It can be shown that this full-rank requirement is satisfied for most scenarios with nondegenerate measurement matrices~$\bA$ or for situations with nonzero measurement noise.
\end{rem}

\subsection{Iterative Variant of PhaseLin}
The authors of \cite{dhifallah2017phase} proposed an iterative scheme that can improve upon the performance of the PhaseMax formulation. The iterative method, called PhaseLamp, first applies PhaseMax with an estimate obtained from a spectral initializer. In each subsequent iteration, the result of the previous iteration is then used as a new initial guess for PhaseMax---this procedure is repeated for a predefined number of iterations or until convergence. Inspired by PhaseLamp, we propose to iteratively apply PhaseLin. The resulting method proceeds as follows.

We start at iteration $t=0$ with a spectral initializer~$\bar\bmx^{(0)}$.
We  run PhaseLin with this initial guess to obtain an estimate of the signal to be recovered $\hat\bmx^{(0)}$. We then take this estimate as a new initial guess, i.e., $\bar\bmx^{(1)}=\hat\bmx^{(0)}$ and re-use PhaseLin to obtain a (hopefully improved) estimate. Concretely, we perform
\begin{align*}
\bar\bmx^{(t+1)} = \hat\bmx^{(t)} = \mathrm{PhaseLin}(\bar\bmx^{(t)}) \quad \text{for} \quad t=0,\ldots,t_\text{max}
\end{align*}
with the final estimate being $\hat\bmx^{(t_\text{max})} $. In our experiments, we simply keep the same covariance matrix $\bC_\bme$  during all iterations. We note  that more sophisticated methods for selecting the error covariance matrix~$\bC_\bme$ on a per-iteration basis may yield improved performance; the design and analysis of such  methods is left for future work.

%

\section{Numerical Results} \label{sec:results}
We now compare the performance of PhaseLin against existing phase retrieval methods for both real and synthetic data.   Algorithm and initializer implementations and experimental setups were provided by PhasePack \cite{chandra2017phasepack}.

\setlength{\textfloatsep}{10pt}
\begin{table}[tp]
\vspace{-0.2cm}
	\centering
	\caption{Relative runtime (RR) and normalized MSE (N-MSE) versus oversampling ratio (OSR) for image recovery as in Figure \ref{fig:TM16}. }
	\begin{tabular}{@{}l|cc|cc@{}}
		\toprule
		Metric  & RR & N-MSE & RR & N-MSE \\
		OSR & \multicolumn{2}{c|}{$M=3N$} &  \multicolumn{2}{c}{$M=9N$}  \\
		\midrule
		PhaseLin-$\complexset$& 1.00&0.3252&1.00&0.2783\\
		WF~\cite{candes2015wirtinger}&0.61 &0.4492&0.13&0.3069\\
		RAF~\cite{wang2017solving} &0.70&0.4769&0.16&0.2946\\ 
		Fienup~\cite{fienup1982phase}&19.1 &0.6070&0.45&0.2899\\
		PhaseMax~\cite{phasemax} &  0.96&0.6254&0.42&0.4872\\
		PhaseLamp~\cite{dhifallah2017phase}&13.3&0.6843&5.31&0.6848\\
		GS~\cite{gerchberg1972practical}&17.9& 0.6036&0.43&0.2899\\
		PhaseLift~\cite{candes2013phaselift}&170  &0.3195&35.0&0.2786\\
		\bottomrule
	\end{tabular}
	\label{tbl:TM16}
	
\end{table}

\subsection{Image Recovery}
We first test the performance of PhaseLin on an image reconstruction task. In this experiment, an image is captured through multiple scattering media, producing phase-less measurements~\cite{metzler2017coherent}; our task is to reconstruct the original image.
In Figure~\ref{fig:TM16}, we show the recovered images for a $16 \times 16$-pixel image taken from the dataset provided in \cite{metzler2017coherent} with $M=3 N$ and $M=9 N$ measurements, respectively. We compare PhaseLin to the 
Wirtinger flow (WF)~\cite{candes2015wirtinger},
reweighed amplitude flow (RAF)~\cite{wang2017solving},
Fienup~\cite{fienup1982phase},
PhaseMax~\cite{bahmani2017phase,goldstein2016phasemax,phasemax},
PhaseLamp~\cite{dhifallah2017phase},
Gerchberg-Saxton (GS)~\cite{gerchberg1972practical},
and PhaseLift~\cite{candes2013phaselift} methods.
For each method, we use the asymptotically-optimal spectral initializer~\cite{mondelli2017fundamental}.
For PhaseLin, we use the complex version  and perform $t_\text{max}=10$ iterations.
\fref{tbl:TM16} provides the associated normalized MSE (N-MSE), which is defined as~\cite{chandra2017phasepack}
\begin{align*}
\textit{N-MSE} = {\min_{\alpha\in\complexset} \|\bmx-\alpha\hat\bmx\|^2}/{\|\bmx\|^2}
\end{align*}
as well as the relative runtime (compared to that of PhaseLin).

We see that for $M=3N$, PhaseLin achieves a lower N-MSE than all other methods except for PhaseLift, which is significantly more complex as it must solve a large semidefinite program. 
Furthermore, for this image dimension, the runtime of PhaseLin is comparable to that of the Wirtinger Flow, RAF, and PhaseMax methods.
For $M=9N$, PhaseLin achieves the lowest N-MSE. However, increasing the number of measurements also increases the complexity of PhaseLin.

\setlength{\textfloatsep}{15pt}
\begin{figure}[tp]
	\centering
	\includegraphics[width=0.95\columnwidth]{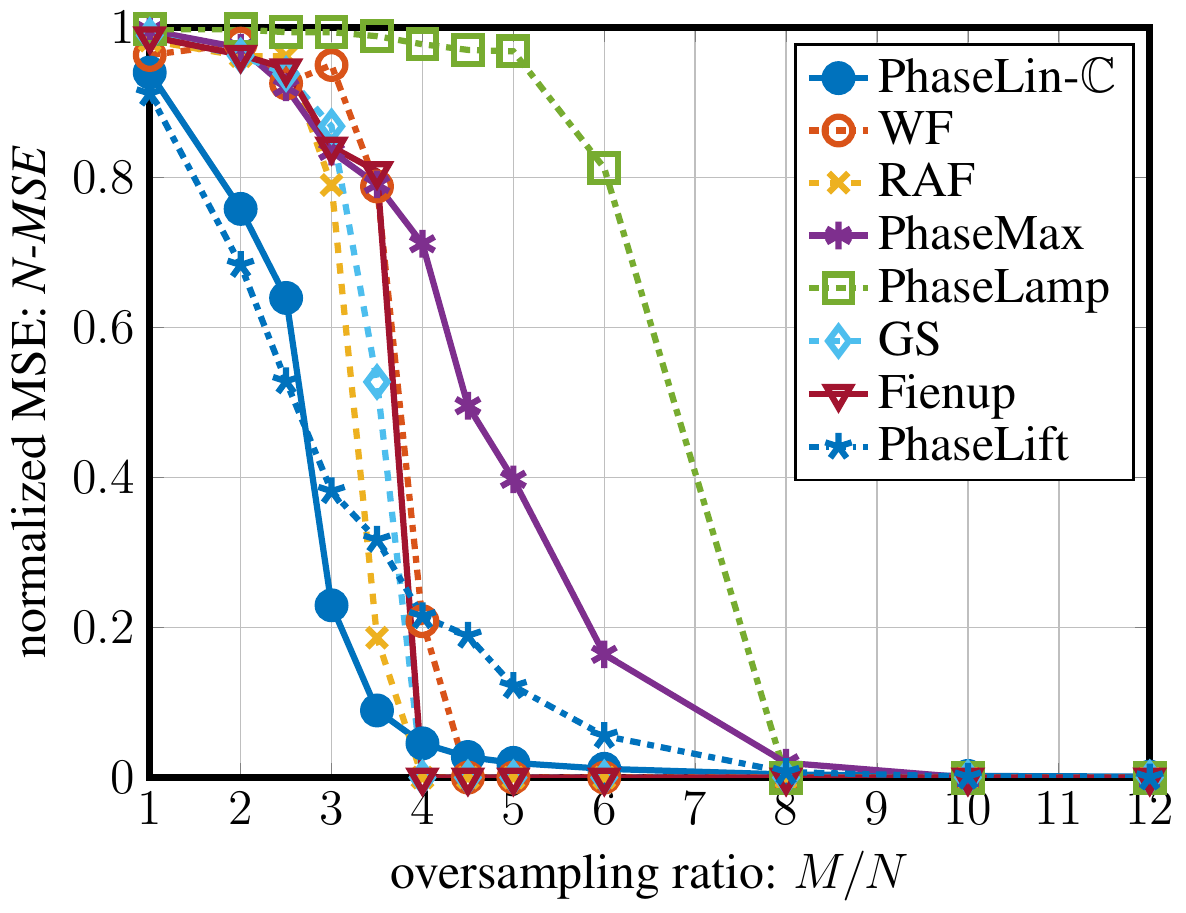}
	\vspace{-0.15cm}
		\caption{Normalized MSE of different phase retrieval methods as a function of the oversampling ratio $M/N$ for Gaussian data with $N=256$ and the same measurement process used in \fref{fig:TM16}. PhaseLin with only 15 iterations outperforms most existing methods in the regime of small oversampling ratios. }
	\label{fig:TM_MN}
\end{figure}

\subsection{Synthetic Data}

To further illustrate the efficacy of PhaseLin, we study its performance for different oversampling ratios $M/N$ using synthetic data in \fref{fig:TM_MN}.
We use the same (empirical) measurement matrix from the above experiment and generate synthetic signals of dimension $256$ from a zero-mean Gaussian distribution as $\bmx \sim \setC \setN(0,2 \bI_{256\times256})$.  We then apply each method to~$10$ randomly generated instances of $\bmx$ and plot the median N-MSE.  
PhaseLin-$\complexset$ with only $15$ iterations outperforms existing methods at small oversampling ratios except PhaseLift (which has much higher computational complexity). 
%
We expect that performance could be improved at higher sampling ratios with more sophisticated strategies to select the error covariance $\bC_\bme$ in each PhaseLin iteration.

\section{Conclusions}
We have proposed PhaseLin, an MSE-optimal linear estimator that recovers signals from magnitude measurements.
PhaseLin requires an initial guess of the true signal that can be obtained from spectral initializers and enables an exact and nonasymptotic analysis of the recovery MSE.
Furthermore, we have demonstrated that PhaseLin performs on par with existing phase retrieval methods using real images and synthetic data when used in an iterative manner. 

There are many avenues for ongoing work. First, analyzing the MSE of the iterative version of PhaseLin is a challenging open research problem. Second, tuning the error covariance in each PhaseLin iteration and further reducing our method's computational complexity is part of ongoing work.


\appendices

\section{Proof of \fref{thm:realPhaseLin}}\label{app:realPhaseLin}
%
Our goal is to derive a linear (or affine) estimate $\hat\bmx^\text{L-MMSE}=\bW\bmy+\bmb$ for the unknown signal $\bmx$ that minimizes the MSE defined in~\fref{eq:MSE}.
The necessary quantities of this linear minimum MSE (L-MMSE) estimator are given by
\begin{align}\label{eq:LMMSE}
 \bW = \bC_{\bmx,\bmy} \bC_\bmy^{-1} \quad \text{and} \quad 
\bmb  = \bar\bmx - \bW\bar\bmy,
\end{align}
with $\bar\bmx=\Ex{}{\bmx}$, $\bar\bmy=\Ex{}{\bmy}$, and 
\begin{align*}
\bC_{\bmx,\bmy} & = \Cov{\bmx,\bmy} =  \Ex{}{(\bmx-\bar\bmx)(\bmy-\bar\bmy)^\Herm} = \Ex{}{\bmx\bmy^\Herm} - \bar\bmx\bar\bmy^\Herm \\
\bC_\bmy & = \Var{\bmy} = \Ex{}{(\bmy-\bar\bmy)(\bmy-\bar\bmy)^\Herm} = \Ex{}{\bmy\bmy^\Herm} - \bar\bmy\bar\bmy^\Herm,
\end{align*}
where we assumed that $\bC_\bmy$ is full rank. 
Our task is to compute the remaining quantities $\bar\bmy$, $\bC_{\bmx,\bmy}$, and $\bC_{\bmy}$.
We will frequently use the following lemma with proof given in \cite[Sec.~3.1]{foldedNormalMoments}.
\begin{lem}\label{lem:foldedvar}
Let  $(u_1,u_2) \sim \setN(\bm\mu,\bSigma)$ be a pair of real-valued jointly Gaussian random variables with covariance matrix 
\begin{align*}	
\bSigma=\left[\begin{array}{ll}\sigma_1^2 & \sigma_{1,2}^2\\\sigma_{1,2}^2&\sigma_2^2\end{array}\right]\!.
\end{align*}
 Then, for $i=1,2$, the pair of random variables $(\nu_1, \nu_2)$ with $\nu_1=u_1^2$ and $\nu_2=u_2^2$ follows the bivariate folded normal distribution with moments
	\begin{align*}
	\bar{\nu_i}&=\Ex{}{u_i^2}=\sigma_i^2+\mu_i^2\\
	[\bC_{\bm\nu}]_{1,2}&=\Ex{}{(\nu_1-\bar{\nu_1})(\nu_2-\bar{\nu_2})} =4\mu_1\mu_2\sigma_{1,2}^2+2\sigma_{1,2}^4\\
	[\bC_{\bm\nu}]_{i,i}&=\Ex{}{(\nu_i-\bar{\nu_i})^2}=2 \sigma_i^4+4\mu_i^2\sigma_i^2.
	\end{align*}
\end{lem}
%
%

Define the vector $\bmz=\bA\bmx+\bmn^z$ to contain the phased measurements, $\bar\bmz=\Ex{}{\bmz}$ its mean vector, and $\bC_\bmz$ its covariance matrix.
Clearly, under \fref{asms:realPhaseLin}, the phased observations in $\bmz$ are jointly Gaussian, with mean $\bar{\bmz}=\bA \bar{\bmx}$ and covariance matrix $\bC_{\bmz}=\bA \bC_{\bme} \bA^\Tran+\bC_{\bmn^z}$.
We are now ready to  compute the missing quantities $\bar\bmy$, $\bC_{\bmx,\bmy}$, and $\bC_{\bmy}$.

\subsubsection{Phased Measurement Mean $\bar\bmy$}
Using \fref{lem:foldedvar}, we have
\begin{align}\label{eq:ymeansq-complex}
\bar{y}_m&=\Ex{}{|z_m|^2+\bmn^y}
=\sigma^2_{m}+|\bar{z}_m|^2+\bar{n}^y_m,
\end{align}
where $\sigma^2_{m}=[\bC_{\bmz}]_{m,m}$.
Hence, the quantity $\bar{\bmy}$ reads
\begin{align*}
\bar\bmy=\diag(\bC_\bmz)+|\bar{\bmz}|^2+\bar{\bmn}^y.
\end{align*} 

\subsubsection{Cross-Covariance Matrix  $\bC_{\bmx,\bmy}$}
To compute the cross-covariance matrix $\bC_{\bmx,\bmy}$, we will use a classical result due to Brillinger given in~\cite[Lem.1]{brillinger1982generalized} that states
\begin{align*}
	\Cov{y_m,x_n} & = \Cov{|z_m|^2,x_n} \\
	& =  \frac{\Cov{|z_m|^2,z_m}}{\Var{z_m}} \Cov{z_m,x_n}.
\end{align*}
By using Stein's lemma \cite{stein}, we have
\begin{align*}
	\frac{\Cov{|z_m|^2,z_m}}{\Var{z_m}} = \Ex{}{2z_m} =2\bar{z}_m.
\end{align*}
We also have 
\begin{align*}
	\Cov{z_m,x_n} &= \Exop[(z_m-\bar{z}_m)(x_n-\bar{x}_n)^*]\\
	&=\bma^\Tran_m\Exop[(\bmx-\bar{\bmx})(x_n-\bar{x}_n)^*]=\bma^\Tran_m[\bC_\bme]_{:,n},
\end{align*}
where $[\bC_\bme]_{:,n}$ corresponds to the $n$th column of $\bC_\bme$ and $\bma^\Tran_m$ to the $m$th row of $\bA$.
Hence, we obtain 
\begin{align} \label{eq:entriescsorrcovariance}
	[\bC_{\bmy,\bmx}]_{m,n} =   2\bar{z}_m \bma^\Tran_m[\bC_\bme]_{:,n},
\end{align}
and $\bC_{\bmx,\bmy}=\bC_{\bmy,\bmx}^\Herm$. 
In compact vector form, the cross-covariance matrix $\bC_{\bmx,\bmy}$ reads
\begin{align*}
\bC_{\bmx,\bmy}=2 \bC_{\bme}\bA^\Tran\diag(\bar{\bmz}).
\end{align*}

\subsubsection{Observation Covariance Matrix $\bC_\bmy$}
The last quantity required is the matrix $\bC_\bmy$. We compute the following quantity:
\begin{align*}
[\bC_\bmy]_{m,m'} & = \Ex{}{(y_m-\bar y_m)(y_m-\bar y_m)^*} \\
&=\Exop[(|z_m|^2+n^y_m-(\Ex{}{|z_m|^2}+\bar{n}^y_m))\\
&\quad \times(|z_{m'}|^2+n^y_{m'}-(\Ex{}{|z_{m'}|^2}+\bar{n}^y_{m'}))^*]\\
&=\Exop[(|z_m|^2-\Ex{}{|z_m|^2})(|z_{m'}|^2-\Ex{}{|z_{m'}|^2})^*] \\
& \quad+[\bC_{\bmn^y}]_{m,m'}.
\end{align*}
Using \fref{lem:foldedvar}, we can compute the above expression as follows.
For $m=m'$, we have
\begin{align*}
[\bC_\bmy]_{m,m}&=
2 \sigma_m^4+4\bar{z}_{m}^2\sigma_m^2,
\end{align*}
where $\sigma_{m}^2=[\bC_{\bmz}]_{m,m}$.
Similarly, for $m \neq m'$, we have
\begin{align*}
[\bC_\bmy]_{m,m'}&=4\bar{z}_{m}\bar{z}_{m'}\sigma_{mm'}^2+2\sigma_{mm'}^4,
\end{align*}
where $\sigma_{mm'}^2=[\bC_{\bmz}]_{m,m'}$.
In summary, the observation covariance matrix $\bC_{\bmy}$ is given by
\begin{align*}
\bC_{\bmy}=
(4\bar{\bmz}\bar{\bmz}^\Tran+2\bC_{\bmz})\odot \bC_{\bmz}+\bC_{\bmn^y}.
\end{align*}

\section{Proof of \fref{thm:complexPhaseLin}}
\label{app:complexPhaseLin}

%
For complex-valued signals, the so-called widely linear minimum MSE (WL-MMSE) estimator~\cite{picinbono1995widely,adali2011complex}, which is of the form $\bW_1 \bmy + \bW_2 \bmy^*$, often provides superior results compared to the standard L-MMSE estimator. 
In our setting, both the WL-MMSE and L-MMSE estimators yield the same results. Hence, for  
\fref{asms:complexPhaseLin}, the linear estimator is simply given by $\hat{\bmx}=\bW \bmy + \bmb$ with $\bW$ and $\bmb$ given in~\fref{eq:LMMSE}.
As a consequence, we only need to compute the following quantities for complex-valued signals:~$\bar\bmy$, $\bC_{\bmx,\bmy}$, and $\bC_{\bmy}$.

Let $\bmz=\bA\bmx+\bmn^z$ contain the phased measurements, $\bar\bmz=\Ex{}{\bmz}$ denote its mean vector, and $\bC_\bmz$ denote its covariance matrix.
Given \fref{asms:complexPhaseLin}, the phased observations in $\bmz$ are jointly complex Gaussian, where we can easily compute the mean $\bar{\bmz}=\bA \bar{\bmx}$ and covariance $\bC_{\bmz}=\bA \bC_{\bme} \bA^{\Herm}+\bC_{\bmn^z}$. 
From the covariance $\bC_\bmz$, we can easily extract the covariance of the real and imaginary parts of $\bmz$ separately as follows: 
\begin{align}\label{eq:CZ_R}
\Ex{}{\bmz_{\mathcal{R}} \bmz_{\mathcal{R}}^*}&\stackrel{(a)}{=}\Ex{}{\bmz_{\mathcal{I}} \bmz_{\mathcal{I}}^*}=1/2\Re\{\bC_\bmz\}=1/2\bC_{\bmz,\mathcal{R}}\\\label{eq:CZ_I}
\Ex{}{\bmz_{\mathcal{I}} \bmz_{\mathcal{R}}^*}&\stackrel{(a)}{=}-\Ex{}{\bmz_{\mathcal{R}} \bmz_{\mathcal{I}}^*}=1/2\Im\{\bC_\bmz\}=1/2\bC_{\bmz,\mathcal{I}}.
\end{align}
Here, (a) follows from the circular symmetry of the complex-valued random variable $\bmx$. We will use these covariances to compute $\bC_{\bmy}$ later in the section.
We are now ready to  compute the missing quantities $\bar\bmy$, $\bC_{\bmx,\bmy}$, and $\bC_{\bmy}$.

\subsubsection{Phased Measurement Mean $\bar\bmy$}
To compute the entries of $\bar\bmy$, we have
\begin{align}\nonumber
\bar{y}_m&=\Ex{}{|z_m|^2+\bmn^y}= \Ex{}{|z_{m,\mathcal{R}}|^2+|z_{m,\mathcal{I}}|^2+n_m^y}\\\nonumber
&\stackrel{\text{(b)}}{=}  |\bar{z}_{m,\mathcal{R}}|^2+\frac{\sigma^2_{m,\mathcal{R}}}{2}+|\bar{z}_{m,\mathcal{I}}|^2+\frac{\sigma^2_{m,\mathcal{R}}}{2}+\bar{n}^y_m\\\label{eq:ymeansq-complex}
&\stackrel{\text{(c)}}{=}|\bar{z}_m|^2+\sigma^2_{m}+\bar{n}^y_m,
\end{align}
where $\sigma^2_{m,\mathcal{R}}=[\bC_{\bmz,\mathcal{R}}]_{m,m}$ and $\sigma^2_{m}=[\bC_{\bmz}]_{m,m}$. Here, (b) follows from \fref{lem:foldedvar}, and (c) follows from the fact that we can conclude $[\bC_{\bmz,\mathcal{R}}]_{m,m}=[\bC_{\bmz}]_{m,m}$ from equations \fref{eq:CZ_R} and~\fref{eq:CZ_I}.
Hence, $\bar{\bmy}$ in vector form reads
\begin{align*}
\bar\bmy=\diag(\bC_\bmz)+|\bar{\bmz}|^2+\bar{\bmn}^y.
\end{align*} 

\subsubsection{Cross-Covariance Matrix $\bC_{\bmx,\bmy}$}
We next compute the individual entries of the  cross-covariance matrix 
\begin{align*}
[\bC_{\bmx,\bmy}]_{n,m}=\Ex{}{(x_n-\bar{x}_n)(y_m-\bar{y}_m)^*}=\Ex{}{x_n y_m^*}-\bar{x}_n\bar{y}_m^*.
\end{align*} 
Let us first focus on the quantity  $\Ex{}{x_ny_m^*}$. We have
%
\begin{align}\nonumber
&\Ex{}{x_n y_m^*}\\\nonumber
&=\Exop \! \left[\! x_n \!\! \left((\sum_{i=1}^{N} \!A_{m,i} x_i+n^z_{m}) (\sum_{j=1}^{N} \!A_{m,j} x_{j}+n^z_{m})^*\!+n^y_m \! \right)^{\!\!\!*}\right]\\\nonumber
&=\Exop\!\left[x_n \sum_{i=1}^{N} A_{m,i} x_i \sum_{j=1}^{N} A_{m,j}^* x_{j}^*\right]+\bar{x}_n [\bC_{\bmn^z}]_{m,m}+\bar{x}_n\bar{n}^y_m\\\nonumber
&=\bar{x}_n \sum_{i=1}^{N} \sum_{j=1}^{N} A_{m,i}  A_{m,j}^* \bar{x}_i \bar{x}_{j}^*\! +\bar{x}_n \sum_{i=1}^{N} \sum_{j=1}^{N} A_{m,i} A_{m,j}^* [\bC_{\bme}]_{i,j} \\\label{eq:Exy}
&+\!\sum_{i=1}^N A_{m,i}\bar{x}_i \sum_{j=1}^{N} A_{m,j}^*  [\bC_{\bme}]_{n,j} +\bar{x}_n [\bC_{\bmn^z}]_{m,m}+\bar{x}_n\bar{n}^y_m.
\end{align}
Next, using \fref{eq:ymeansq-complex}, we obtain
\begin{align}\nonumber
\bar{x}_n\bar{y}_m^*&=\bar{x}_n (\sigma^2_{n}+|\bar{z}_m|^2+\bar{n}^y_m)\\\nonumber
&=\bar{x}_n ([\bA \bC_{\bme}\bA^\Herm+\bC_{\bmn^z}]_{m,m}+| \bma_m^\Herm \bar{\bmx}|^2+\bar{n}^y_m)\\\nonumber
&=\bar{x}_n \sum_{i=1}^{N} \sum_{j=1}^{N} A_{m,i} A_{m,j}^* [\bC_{\bme}]_{i,j}+ \bar{x}_n [\bC_{\bmn^z}]_{m,m} \\\label{eq:Exbarybar}
&+ \bar{x}_n \sum_{i=1}^{N} \sum_{j=1}^{N} A_{m,i}  A_{m,j}^* \bar{x}_i \bar{x}_{j}^*+\bar{x}_n\bar{n}^y_m.
\end{align}
Hence, from \fref{eq:Exy} and \fref{eq:Exbarybar}, we finally get
\begin{align*}
[\bC_{\bmx,\bmy}]_{n,m} = \sum_{i=1}^N A_{m,i}\bar{x}_i \sum_{j=1}^{N} A_{m,j}^*  [\bC_{\bme}]_{n,j},
\end{align*}
which, in compact form, reads 
$\bC_{\bmx,\bmy}= \bC_{\bme}\bA^\Herm \diag(\bar{\bmz})$.
%

\subsubsection{Observation Covariance Matrix $\bC_\bmy$}
To obtain the entries of the matrix $\bC_\bmy$, we first calculate 
\begin{align*}
[\bC_\bmy]_{m,m'} & = \Ex{}{(y_m-\bar y_m)(y_m-\bar y_m)^*} \\
&=\Exop[(|z_m|^2+n^y_m-(\Ex{}{|z_m|^2}+\bar{n}^y_m))\\
	&\; \times(|z_{m'}|^2+n^y_{m'}-(\Ex{}{|z_{m'}|^2}+\bar{n}^y_{m'}))^*]\\
	&=\Exop[(|z_{m,\mathcal{R}}|^2\! -\Ex{}{|z_{m,\mathcal{R}}|^2} \!+\!|z_{m,\mathcal{I}}|^2 \!-\Ex{}{|z_{m,\mathcal{I}}|^2})\\
	& \; \times \! (|z_{{m'},\mathcal{R}}|^2\!\!-\!\Ex{}{|z_{{m'},\mathcal{R}}|^2}\!\!+\!|z_{{m'},\mathcal{I}}|^2\!\!-\!\Ex{}{|z_{{m'},\mathcal{I}}|^2})^*]\\
	&\; +[\bC_{\bmn^y}]_{m,m'}.
\end{align*}
Using the expressions in \fref{eq:CZ_R} and \fref{eq:CZ_I} together with  \fref{lem:foldedvar}, we can compute the above expressions.
For $m=m'$, we have
\begin{align*}
[\bC_\bmy]_{m,m}&=2 {\frac{\sigma_{m,\mathcal{R}}^4}{4}}+4\bar{z}_{m,\mathcal{R}}^2{\frac{\sigma_{m,\mathcal{R}}^2}{2}} +2 {\frac{\sigma_{m,\mathcal{R}}^4}{4}}+4\bar{z}_{m,\mathcal{I}}^2{\frac{\sigma_{m,\mathcal{R}}^2}{2}}\\
&\;+4\bar{z}_{m,\mathcal{I}}\bar{z}_{m,\mathcal{R}}{\frac{\sigma_{m,\mathcal{I}}^2}{2}}+2{\frac{\sigma_{m,\mathcal{I}}^4}{4}}+4\bar{z}_{m,\mathcal{R}}\bar{z}_{m,\mathcal{I}}{\frac{\sigma_{m,\mathcal{I}}^2}{2}}\\
&\;+2{\frac{\sigma_{m,\mathcal{I}}^4}{4}}+[\bC_{\bmn^y}]_{m,m},
\end{align*}
where $\sigma_{m,\mathcal{R}}^2=[\bC_{\bmz,\mathcal{R}}]_{m,m}$ and $\sigma_{m,\mathcal{I}}^2=[\bC_{\bmz,\mathcal{I}}]_{m,m}$. From \fref{eq:CZ_I} we see that $\sigma_{m,\mathcal{I}}^2=0$.
Hence, $[\bC_\bmy]_{m,m}$ can be written as
\begin{align*}
[\bC_\bmy]_{m,m}&=\sigma_{m,\mathcal{R}}^4+2(\bar{z}_{m,\mathcal{R}}^2+\bar{z}_{m,\mathcal{I}}^2)\sigma_{m,\mathcal{R}}^2+[\bC_{\bmn^y}]_{m,m}.
\end{align*}
Similarly, for $m \neq m'$ we have
\begin{align*}
[\bC_\bmy]_{m,{m'}}&=4\bar{z}_{m,\mathcal{R}}\bar{z}_{{m'},\mathcal{R}}{\frac{\sigma_{m{m'},\mathcal{R}}^2}{2}}+2{\frac{\sigma_{m{m'},\mathcal{R}}^4}{4}}\\
&+4\bar{z}_{m,\mathcal{I}}\bar{z}_{{m'},\mathcal{I}}{\frac{\sigma_{m{m'},\mathcal{R}}^2}{2}}+2{\frac{\sigma_{m{m'},\mathcal{R}}^4}{4}}\\
&-4\bar{z}_{m,\mathcal{R}}\bar{z}_{{m'},\mathcal{I}}{\frac{\sigma_{m{m'},\mathcal{I}}^2}{2}}+2{\frac{\sigma_{m{m'},\mathcal{I}}^4}{4}}\\
&+4\bar{z}_{m,\mathcal{I}}\bar{z}_{{m'},\mathcal{R}}{\frac{\sigma_{m{m'},\mathcal{I}}^2}{2}}+2{\frac{\sigma_{m{m'},\mathcal{I}}^4}{4}}\\
&+[\bC_{\bmn^y}]_{m,{m'}},
\end{align*}
where $\sigma_{m{m'},\mathcal{R}}^2=[\bC_{\bmz,\mathcal{R}}]_{m,{m'}}$ and $\sigma_{m{m'},\mathcal{I}}^2=[\bC_{\bmz,\mathcal{I}}]_{m,{m'}}$. 
As a result, the observation covariance matrix  $\bC_{\bmy}$ is given by
\begin{align*}
\bC_{\bmy}=2\Re\left\{\left(\bar{\bmz}\bar{\bmz}^\Herm\right) \odot \bC_{\bmz} ^* \right\}+\bC_{\bmz} \odot \bC_{\bmz}^*+\bC_{\bmn^y}.
\end{align*}

\section{Proof of \fref{lem:MSE}}
\label{app:MSE}
The MSE of the PhaseLin estimator is given in \fref{eq:MSE}.
If $\bC_\bmy$ is full rank, then $\hat\bmx = \bC_{\bmx,\bmy} \bC_{\bmy}^{-1} (\bmy-\bar\bmy) + \bar\bmx$.
Inserting this PhaseLin estimator in the MSE expression leads to 
\begin{align*}
\textit{MSE} & = \Ex{\,\bme,\bmn^z,\bmn^y}{\|\bC_{\bmx,\bmy} \bC_{\bmy}^{-1} (\bmy-\bar\bmy) + \bar\bmx-\bmx\|^2_2} \\
& = \tr{(\Ex{\,\bme,\bmn^z,\bmn^y}{\bC_{\bmx,\bmy} \bC_{\bmy}^{-1} (\bmy-\bar\bmy) (\bmy-\bar\bmy)^\Herm \bC_{\bmy}^{-1} \bC_{\bmx,\bmy}^\Herm})} \\
& \quad - \tr{(\Ex{\,\bme,\bmn^z,\bmn^y}{\bC_{\bmx,\bmy} \bC_{\bmy}^{-1} (\bmy-\bar\bmy) (\bmx-\bar\bmx)^\Herm})} \\
& \quad - \tr{(\Ex{\,\bme,\bmn^z,\bmn^y}{(\bmx-\bar\bmx) (\bmy-\bar\bmy)^\Herm \bC_{\bmy}^{-1} \bC_{\bmx,\bmy}^\Herm})} \\
& \quad + \tr{(\Ex{\,\bme}{(\bmx-\bar\bmx)(\bmx-\bar\bmx)^\Herm})} \\
& = \tr{(\bC_{\bmx,\bmy} \bC_{\bmy}^{-1} \bC_{\bmx,\bmy}^\Herm)} - \tr{(\bC_{\bmx,\bmy} \bC_{\bmy}^{-1} \bC_{\bmy,\bmx})} \\ 
& \quad - \tr{(\bC_{\bmx,\bmy} \bC_{\bmy}^{-1} \bC_{\bmx,\bmy}^\Herm)} + \tr{(\bC_{\bme})} \\
& = \tr{(\bC_{\bme} - \bC_{\bmx,\bmy} \bC_{\bmy}^{-1} \bC_{\bmy,\bmx})},
\end{align*}
where we have used the fact that $\bC_{\bmy,\bmx} = \bC_{\bmx,\bmy}^\Herm$. 





\balance

\bibliographystyle{IEEEtran} 
\bibliography{confs-jrnls,publishers,PhaseLin}


\end{document}